\documentclass[12pt]{article}
\usepackage{amsmath}
\usepackage{epsfig}
\usepackage{amssymb,amsfonts}
\usepackage[all]{xy}

\advance\voffset by -2.0cm
\advance\hoffset by -1.25cm
\def\lto{\longrightarrow}
\def\Coker{\operatorname{Coker}\,}

\def\BC{{\mathbb C}}
\def\be{\begin{equation}}
\def\ee{\end{equation}}
\def\ba{\begin{eqnarray}}
\def\ea{\end{eqnarray}}

\renewcommand{\H}{{\cal H}}

\newcommand{\BZ}{\mathbb{Z}}

\newcommand{\Real}{\mathbb{R}}

\newcommand{\Zop}{\mathbb{Z}}

\begin{document}

\vspace*{-1.5cm}
\thispagestyle{empty}
\begin{flushright}
hep-th/0506208
\end{flushright}
\vspace*{2.5cm}

\begin{center}
{\Large 
{\bf The matrix factorisations of the D-model}}
\vspace{2.0cm}

{\large Ilka Brunner}%
\footnote{{\tt E-mail: brunner@itp.phys.ethz.ch}} 
{\large and} {\large Matthias R.\ Gaberdiel}%
\footnote{{\tt E-mail: gaberdiel@itp.phys.ethz.ch}} 
\vspace*{0.5cm}

Institut f\"ur Theoretische Physik, ETH-H\"onggerberg\\
8093 Z\"urich, Switzerland\\
\vspace*{2cm}

{\bf Abstract}
\end{center}

\noindent The fundamental matrix factorisations of the D-model
superpotential are found and identified with the boundary states of
the corresponding conformal field theory. The analysis is performed
for both GSO-projections. We also comment on the relation of
this analysis to the theory of surface singularities and their 
orbifold description.

\newpage
\renewcommand{\theequation}{\arabic{section}.\arabic{equation}}


\section{Introduction}

Models with $N=(2,2)$ worldsheet supersymmetry play a major role in
various applications of string theory, most notably in the context
of Calabi-Yau compactifications, mirror symmetry and the construction
of four-dimensional string vacua. The simplest non-trivial theories
with superconformal symmetry are the $N=2$ minimal models which have
an ADE classification. These models play a central role in the
construction of Gepner models that describe certain Calabi-Yau
compactifications at specific points in their moduli space. In
addition to their realisation as abstract conformal field theories,
the $N=2$ minimal models also have a description as Landau Ginzburg
theories. For the case of the D-model that shall mainly concern us in
this paper, the relevant superpotentials are   
\be
W_D=x^{\frac{n+1}{2}} - xy^2 \,, \qquad
W_{D'} = x^{\frac{n+1}{2}} - xy^2 - z^2 \,.
\ee
The two different choices correspond to the two possible
GSO-projections that can be imposed in conformal field theory.

In this paper, we perform a complete analysis of the B-type branes in
both of these theories. These can be studied from two points of view:
on the one hand, they have a description in terms of boundary states  
in rational conformal field theory. On the other hand, according
to a proposal in unpublished work of Kontsevich, B-type
branes in Landau Ginzburg models can be characterised in terms of
matrix factorisations of the superpotential. This proposal was discussed 
in the physics literature in \cite{KL1,BHLS,KL2,KL3,L,HL,HW,HWs}.
One would thus expect that the boundary state analysis in
conformal field theory should agree with that of the
D-branes in the Landau Ginzburg theory. This was for
example seen to be the case for the A-type models in \cite{BHLS,KL3}.   

In the following we shall compare in detail the boundary states of the
D-type minimal models with the matrix factorisations of the
corresponding Landau Ginzburg models. Furthermore, we shall relate the
matrix factorisations of the Landau Ginzburg superpotential $W$ to
matrix factorisations that appear in the study of singularity theory.
Surface singularities can formally be related to Landau Ginzburg
theories by studying the locus $W=0$ as a hypersurface in $\BC^3$. The
matrix factorisations encode then the geometrical details of the
blow-ups necessary to resolve the surface singularity. Interpreted on
the singular variety, they also correspond to the elements of Orlov's
category $D_{Sg}$ \cite{orlov}. Our analysis suggests that the
Landau-Ginzburg theory, despite having smaller central charge,
captures some of the physical properties of the D-branes on these
surface singularities that become massless at the singular point.

The paper is organised as follows. In section 2 we review the relation
between matrix factorisations and boundary states for the A-type
minimal models. As explained in \cite{KL3}, adding a square to the
Landau Ginzburg superpotential corresponds to a change in the GSO
projection in conformal field theory. In section 3 we construct all
B-type boundary states in the D-model for both possible GSO
projections. Since the D-model is a rational conformal field theory
(with respect to the $N=2$ superconformal algebra), the boundary
states we construct generate {\it all} superconformal boundary states.
In section 4 we discuss the matrix factorisations of the corresponding
Landau Ginzburg models, completing the list of factorisations provided
in \cite{KL3}. Once these additional factorisations are considered, we
obtain complete agreement with the conformal field theory description.
On the other hand, since the latter description is complete (in the
above sense), we can conclude that the matrix factorisations we have
found are all the fundamental matrix factorisations for these
superpotentials. This is to say, any matrix factorisation must be the
direct sum of these fundamental factorisations. Our analysis also
confirms the conjecture of \cite{KL3} regarding the relation between
the different GSO-projections in conformal field theory and the
addition of a square to the Landau Ginzburg model. Finally, in
section~5, we investigate the relation between D-branes on singular
surfaces, the corresponding Landau Ginzburg theories and conformal
field theory. We also comment on the description of the surface
singularity in terms of the singular quotient $\BC^2/\Gamma$, where
$\Gamma$ is a subgroup of $SU(2)$. This allows us to use the methods
of \cite{DM,JM} to describe D-branes on those orbifolds in terms of
representations of $\Gamma$. We find complete agreement with the
results from the matrix factorisation point of view, confirming yet
again that we have found all D-branes.


\section{Matrix factorisations}
\setcounter{equation}{0}

Let us briefly review the description of D-branes in terms of matrix
factorisations. This approach was proposed in unpublished form by
Kontsevich, and the physical interpretation of it was given in
\cite{KL1,BHLS,KL2,KL3,L,HL}; for a good review of this material see
for example \cite{HWs}. 

According to Kontsevich's proposal, D-branes in Landau Ginzburg models
correspond to matrix factorisations of the superpotential $W(x_i)$,
\be\label{factorize}
d_0\, d_1=d_1\, d_0 = W\, {\bf 1}\,,
\ee
where $d_0$ and $d_1$ are $r\times r$ matrices with polynomial
entries in the $x_i$. To a matrix factorisation of this form,  
one can then associate the boundary BRST operator $Q$ of the form  
\be\label{Q}
Q=\left( \begin{array}{cc} 0 & d_1 \\ d_0 & 0 \end{array} \right) \,.
\ee
The spectrum of open string operators consists of polynomial matrices
of the same size. It is naturally $\BZ_2$ graded, where the bosons
$\phi$ correspond to block-diagonal matrices, while the fermions $t$
are off-diagonal:    
\be
\phi = \left( \begin{array}{cc} \phi_0 & 0 \\ 
0 & \phi_1 \end{array} \right)\,, \qquad
\qquad
t= \left( \begin{array}{cc} 0 & t_1 \\ t_0 & 0 \end{array} \right) \,.
\ee
To obtain the physical spectrum between two such branes, one restricts
to the degree $0$ operators satisfying $D\phi=[Q,\phi]=0$, and
identifies operators that differ only by BRST-exact terms. To
calculate this cohomology, we can follow the strategy of \cite{KL3}. A 
BRST invariant boson has to satisfy
\be
d_1 \, \phi_1 = \phi_0\, d_1\,, \qquad \hbox{and} \qquad
d_0 \, \phi_0 = \phi_1\, d_0 \,,
\ee
where $\phi_0$ and $\phi_1$ are $r\times r$ matrices. This equation can  
be solved for $\phi_0$ as 
\be\label{phi0}
\phi_0 = \frac{1}{W} \, d_1\, \phi_1 \, d_0 \,,
\ee
provided that the matrix on the right hand side is divisible by $W$. 
After moding out $Q$-exact terms, the bosonic cohomology can
therefore be described using $\phi_1$ only. The BRST trivial
$\tilde\phi_1$ are derivatives of fermions, and hence take the form 
\be\label{BRSTex}
\tilde\phi_1 = (Dt)_1=t_0d_1+d_0t_1 \,.
\ee
\smallskip

\noindent For the fermions, the condition for BRST invariance is 
\be
t_0\, d_1 + d_0 \,t_1 = 0 = d_1\, t_0 + t_1\, d_0 \,,
\ee
which can be solved for $t_1$
\be\label{t1}
t_1 = - \frac{1}{W} \, d_1 \, t_0 \, d_1\,, 
\ee
resulting again in a divisibility condition. An invariant fermion 
$(\tilde{t}_0, \tilde{t}_1)$ is BRST trivial if $\tilde{t}_0$ can be
written as 
\be
\tilde{t}_0=-\phi_1\, d_0 + d_0\, \phi_0\,,
\ee
for a boson $(\phi_0, \phi_1)$.
\smallskip

It is easy to see that the spectrum of two factorisations $Q$ and
$\hat{Q}$ is the same, if 
\begin{equation}\label{equiv}
Q = U \, \hat{Q}\, U^{-1}\,, \qquad
U = \left( \begin{array}{cc} A & 0 \\ 
0 & B \end{array} \right)\,,
\end{equation}
provided that $U$ and its inverse, $U^{-1}$, are polynomial
matrices. We therefore identify two such factorisations \cite{HLL1}. 

For a given factorisation $Q$, we call the factorisation where the
roles of $d_0$ and $d_1$ are reversed the {\it reverse factorisation}
$Q^r$. It is clear from the above discussion that the bosonic spectrum
between two factorisations $Q_1$ and $Q_2$ and the fermionic spectrum
between $Q_1$ and $Q_2^r$ coincide. 

To make contact with conformal field theory, one has to make 
sure that the $U(1)$ $R$-symmetry that becomes the
$U(1)$ current symmetry of the $N=2$ superconformal algebra in the IR 
is preserved \cite{HW}.  This means that one has to restrict to
homogeneous superpotentials and specify a consistent assignment of
R-charge in the boundary theory. By (\ref{factorize}) and (\ref{Q})
this implies that $Q$ should have charge one 
\be
e^{i\lambda R} \ Q(e^{i\lambda q_i} x_i) \ e^{-i\lambda R} 
= e^{i\lambda} \ Q \,.
\ee
Then one can assign R-charge to the boundary operators by
\cite{HW} 
\be
e^{i\lambda R} \ \phi(e^{i\lambda q_i} x_i) \ e^{-i\lambda R} = 
e^{i\lambda q} \ \phi \, .
\ee

This description of D-branes should then have a direct correspondence in
conformal field theory. In particular, the different factorisations
(up to the aforementioned equivalence) should correspond to the
different B-type boundary states in conformal field theory. The
physical boson spectrum described above corresponds then to the
topological open string spectrum from the point of view of conformal
field theory. The reverse factorisation of a given factorisation
corresponds to the anti-brane of the given brane; thus the physical
fermion spectrum in the above description corresponds to the
topological open string spectrum between a brane and an anti-brane.  

In order to illustrate these concepts and set up notation for the rest
of the paper, we briefly review the case of the A-type minimal models.

\subsection{A-type minimal models}

In \cite{BHLS,KL3} A-type minimal models were studied from the matrix
point of view. For the potential 
\be\label{ASP}
W_A=x^n
\ee
it can be shown that a class of inequivalent matrix factorisations that
generate all factorisations are described by the simple rank $1$
factorisations $W=x^{l} x^{n-l}$, where $l=1,\dots, n-1$. The 
corresponding BRST operator is 
\be\label{Ql}
Q_l=\left( \begin{array}{cc} 0 & x^l \\ x^{n-l} & 0 \end{array} \right) \,.
\ee
The factorisations $Q_l$ and $Q_{n-l}$ are related to one another by 
the exchange of $d_0$ and $d_1$; they are therefore reverse
factorisations of one another. 

The corresponding conformal field theory is described by a single
$N=2$ minimal model with $n=k+2$. (Our conventions are chosen as in
\cite{BG1}.) The spectrum of this theory is (after GSO-projection) 
\be\label{single}
\H_A = \bigoplus_{[l,m,s]} \,
\left( \H_{[l,m,s]} \otimes \bar\H_{[l,m,-s]} \right) \,.
\ee
The GSO projection chosen here is the analogue of the Type 0A projection.
B-type branes are characterised by the gluing conditions 
\ba
\left(L_n - \bar{L}_{-n} \right) |\!| B \rangle\!\rangle & = & 0
\nonumber \\
\left(J_n + \bar{J}_{-n} \right) |\!| B \rangle\!\rangle & = & 0
\label{gluing} \\
\left(G^\pm_r + i\, \eta\, \bar{G}^\pm_{-r} \right) 
|\!| B \rangle\!\rangle & = & 0 \,. \nonumber
\ea
The corresponding B-type boundary states were constructed some time
ago (see for example \cite{MMS}), and are explicitly given as  
\be
|\!| L,S \rangle\!\rangle = 
\sqrt{k+2} \,
\sum_{l+s\in 2\Zop} 
\frac{S_{L0S,l0s}}{\sqrt{S_{l0s,000}}} \, 
| [l,0,s]\rangle\!\rangle \,.
\ee
Here $L=0,1,\ldots,k$ and $S=0,1,2,3$. The boundary states with $S$
even (odd) satisfy the gluing conditions with $\eta=+1$ ($\eta=-1$); 
in the following we shall restrict ourselves to the case $\eta=+1$,
and thus to even $S$.\footnote{The D-branes corresponding to $\eta=-1$
or $S$ odd preserve a different supercharge at the boundary. The
branes that are described by the different matrix factorisations
however always preserve the {\it same} supercharge at the boundary.} 
We also note that 
\be\label{Aantibrane}
|\!| L,S \rangle \!\rangle = |\!| k-L,S+2\rangle\!\rangle
\ee
and thus there are only $k+1$ different boundary states with $\eta=+1$
(and $k+1$ different boundary states with $\eta=-1$). These boundary
states therefore account for all the $N=2$ Ishibashi states. Finally
we note that $|\!| L,S\rangle\!\rangle$ and  
$|\!| L,S+2\rangle\!\rangle$ are anti-branes of one another (since
they differ by a sign in the coupling to the RR sector states). 

It is then suggestive to identify the matrix factorisation
corresponding to $Q_l$ (\ref{Ql}) with the boundary state with label 
$L=l-1$ and $S=0$. The equivalence (\ref{Aantibrane}) mirrors the 
factorisation reversal map $l\mapsto n-l$ on the matrix
side. Restricting without loss of generality to the range $l\leq n/2$
one can then easily check, both in conformal field theory and the
matrix language, that there are $l$ topological states in the open
string spectrum on each of these branes (they correspond to the
`bosons' from the matrix point of view), and $l$ topological states in
the open string spectrum between brane and anti-brane (the `fermions'
in the matrix description). 
\bigskip

{}From the conformal field theory perspective it is immediately 
clear that there is another closely related theory, where one uses a
type 0B like GSO projection instead of the GSO projection discussed
above. The relevant Hilbert space is
\be\label{single'}
\H_{A'} = \bigoplus_{[l,m,s]} \,
\left( \H_{[l,m,s]} \otimes \bar\H_{[l,m,s]} \right) \,.
\ee
It was shown in \cite{KL3} that from the matrix point of view this
theory is described by the superpotential 
\be
W_{A'}=x^n-y^2 \,.
\ee
Since the D-models that we shall discuss in this paper are
closely related to this theory, let us briefly summarise the results
of \cite{KL3} on factorisations and their relation to boundary
states. The superpotential $W_{A'}$ has two classes of factorisations:
the first class is given by   
\be\label{Afactor}
d_0= 
\left(\begin{array}{cc} x^l & y \\ -y & -x^{n-l} \end{array}
\right)\,,  
\qquad
d_1= \left(\begin{array}{cc} x^{n-l} & y \\ -y & -x^{l} \end{array}
\right) \,. 
\ee
One can easily see that the factorisation corresponding to $l$ and
$n-l$ are equivalent in the sense of (\ref{equiv}); in the following
we shall therefore always take $l\leq n/2$. Since these are 
reverse factorisations of one another, the corresponding branes
have to be their own anti-branes.

These factorisations are just a special class of the tensor product
factorisations discussed in \cite{ADD}. In fact, they consist of one   
factor coming from the potential $x^n$, and another one from the
`trivial' factor $y^2$ whose physical significance has been
discussed in \cite{KL1}. Accordingly, the open string spectrum
inherits the tensor product structure: there are $2l$ bosons of the
type 
\be\label{Abosons}
a_i=\left( \begin{array}{cc} x^i & 0 \\ 0 & x^i \end{array} \right)\,, 
\qquad
a_{l+i}=\left( \begin{array}{cc} 0 & x^i \\ x^{n-2l+i} & 0 \end{array} 
\right)\,, \quad i=0,\dots , l-1\,, 
\ee
where the $2\times 2$ matrices $a_j$ describe $\phi_1$, and $\phi_0$
is then determined by (\ref{phi0}). Likewise, there are $2l$ fermions
\be
\eta_i=\left( \begin{array}{cc}0 & x^i \\  x^i & 0 \end{array}
\right)\,, \qquad
\eta_{l+i}=\left( \begin{array}{cc} x^{i} & 0 \\ 0 & x^{n-2l+i} 
\end{array} \right)\,, \quad i=0,\dots , l-1 \,,
\ee
where the $2\times 2$ matrices $\eta_j$ describe $t_0$, and $t_1$ is
then determined by (\ref{t1}). 
In the case that $n$ is even, the factorisation $l=n/2$ is not
irreducible. In fact, it is equivalent (in the sense of (\ref{equiv})) 
to the direct sum of the rank $1$ factorisation
\be\label{prim}
d_0= (x^{\frac{n}{2}} + y) \,, \qquad d_1=(x^{\frac{n}{2}} - y) \,,
\ee
and its reverse. The physical open string spectrum of these
rank $1$ factorisations consists of $n/2$ bosons (but no fermions);
correspondingly there are $n/2$ fermions (but no bosons) between the
factorisation (\ref{prim}) and its reverse.
\medskip

In conformal field theory, the rank $2$ factorisations reviewed above 
correspond to the boundary states \cite{KL3} 
\be
| \!|L,S\rangle\!\rangle =(2k+4)^{1/4} \sum_{l\, {\rm even}} 
\sum_{\nu=0,1} 
\frac{S_{Ll}}{\sqrt{S_{0l}}} \, 
e^{-i\pi \nu S}\, 
|[l,0,2\nu]\rangle\!\rangle  \, . 
\ee
Here $S$ is only defined modulo $2$, and 
$| \!|L,S\rangle\!\rangle = | \!|k-L,S\rangle\!\rangle$. 
As before, the identification
relates the factorisation (\ref{Afactor}) to the boundary state with
$L=l-1$ and $S=0$. (Recall that $n=k+2$.) These boundary states do not
couple to RR states and are therefore equivalent to their own
anti-branes. (This is simply the statement that $S$ is only defined
modulo $2$.)

If $k$ is even, then the $L=k/2$ boundary state contains two vacua in
its open string spectrum and thus can be further resolved. The
explicit form of the two resolved boundary states is \cite{MMS}
\be
|\!|k/2,S\rangle\!\rangle_{\rm res}= \frac{1}{2} \left(
|\!|k/2,S\rangle\!\rangle + \sqrt{k+2} 
\sum_{s=\pm 1} e^{-\frac{\pi i Ss}{2}} 
|[k/2,(k+2)/2,s]\rangle\!\rangle \right)\,, 
\ee
where $S$ is now defined mod $4$, and
$|\!|k/2,S\rangle\!\rangle_{\rm res}$ and 
$|\!|k/2,S+2\rangle\!\rangle_{\rm res}$ are anti-branes of one
another. The two branes with $S=0$ and $S=2$ then correspond to the
matrix factorisation (\ref{prim}) and its reverse, respectively.
\bigskip

We shall now construct the boundary states of the various
GSO-projections of the D-model; in section~4 we shall discuss the
corresponding matrix factorisations and identify the two
descriptions. 

\section{Boundary states for the D-model}
\setcounter{equation}{0}

The D-model is only defined if the level $k$ is even. As before for
the case of the A-type minimal model there are two possible
GSO-projections one can consider. We begin by discussing the theory
that is analogous to type 0B. Its spectrum depends on whether $k/2$ is
even or odd,  {\it i.e.} on whether $k$ is divisible by $4$ or not. In
the former case, the spectrum is  
\be\label{neven}
\H = \bigoplus_{[l,m,s],\ l\ {\rm even}}
\Bigl( \left(\H_{[l,m,s]} \otimes \bar\H_{[l,m,s]}\right)  
\oplus 
\left(\H_{[l,m,s]} \otimes \bar\H_{[k-l,m,s]}\right) 
\Bigr) \,.
\ee
On the other hand, if $k/2$ is odd the spectrum is 
\be\label{nodd}
\H = \bigoplus_{[l,m,s],\ l\ {\rm even}}
\Bigl( \left(\H_{[l,m,s]} \otimes \bar\H_{[l,m,s]}\right)   \Bigr) 
\, \oplus \,
\bigoplus_{[l,m,s],\ l\ {\rm odd}} 
\Bigl( 
\left(\H_{[l,m,s]} \otimes \bar\H_{[k-l,m,s]}\right) 
\Bigr) \,.
\ee
We are interested in the B-type boundary states of this theory. In our
conventions, the B-type Ishibashi states are characterised by the
gluing conditions (\ref{gluing}). The structure of these Ishibashi
states (and thus the corresponding boundary states) depends on the two
cases above, and we shall therefore discuss them in turn.

\subsection{The case $k/2$ odd}

For $k/2$ odd, the theory (\ref{nodd}) has the Ishibashi states 
\be
|[l,0,s]\rangle\!\rangle \in \H_{[l,0,s]}\otimes \bar\H_{[l,0,s]} \,,
\ee
where $l$ and $s$ are both even, and the Ishibashi states 
\be
|[l,(k+2)/2,s]\rangle\!\rangle \in
\H_{[l,\frac{k+2}{2},s]}\otimes \bar\H_{[k-l,\frac{k+2}{2},s]}
\ee
where $l$ and $s$ are both odd. In total there are therefore $2(k+1)$
Ishibashi states of the coset theory; they give rise to $(k+1)$
Ishibashi states of the $N=2$ theory with one spin structure
$\eta=+1$, and  $(k+1)$ Ishibashi states of the $N=2$ algebra with
the other $\eta=-1$. 

\noindent The corresponding boundary states are described by 
\begin{eqnarray}
|\!| L,M,S \rangle\!\rangle & = & \sqrt{\frac{2k+4}{2}} \, 
\Biggl( \sum_{l,s\  {\rm even}} 
\frac{S_{LMS,l0s}}{\sqrt{S_{000,l0s}}}\, |[l,0,s]\rangle\!\rangle  
\nonumber \\
& & \qquad \qquad \qquad \qquad 
+  \sum_{l,s\  {\rm odd}} 
\frac{S_{LMS,l\frac{k+2}{2}s}}{\sqrt{S_{000,l\frac{k+2}{2}s}}}\, 
|[l,(k+2)/2,s]\rangle\!\rangle  \Biggr) \,. \label{boundnodd}
\end{eqnarray}
Here $L+M+S$ is even, and we have the identifications
\begin{eqnarray}
|\!|L,M,S\rangle\!\rangle & = & |\!| k-L,M+k+2,S+2 \rangle \!\rangle 
= |\!| k-L,M,S+2 \rangle \!\rangle \nonumber \\
& = & |\!| L,M+4,S \rangle \!\rangle = 
|\!| L,M+2,S+2 \rangle \!\rangle \,.
\end{eqnarray}
For fixed spin structure $\eta=+1$ (corresponding to $S=0,2$), we have
for every value of $L$ only two possible values of $M$, namely $M=0,2$
or $M=1,3$. Furthermore, because of the last identification, of the
four choices ($S=0,2$, $M=0,2$) or ($S=0,2$, $M=1,3$) only two
describe different boundary states. For $L\neq k/2$, the pair of
boundary states with  $L$ and $k-L$ account therefore each for two
different boundary states giving in total $k$ different boundary
states. For $L=k/2$ on the other hand, it is easy to see that the
second sum in (\ref{boundnodd}) vanishes, and thus  
$|\!|k/2,+1,0\rangle\!\rangle = |\!|k/2,-1,0\rangle\!\rangle$, thus
giving only one additional boundary state. In total the above
construction therefore gives $k+1$ different boundary states of fixed
spin structure, and thus accounts for all the Ishibashi states.

These boundary states satisfy the Cardy condition since their overlap
equals 
\begin{eqnarray}\label{noddspec}
&& \langle\!\langle L_1,M_1,S_1 |\!|  
q^{L_0 + \bar{L}_0 - \frac{c}{12}} \,
|\!| L_2,M_2,S_2 \rangle \!\rangle 
=  \sum_{[l,m,s]} \, \chi_{(l,m,s)}(\tilde{q}) \, 
\delta^{(2)}(S_2+s-S_1) \times \nonumber \\
& & \qquad \qquad \times
\Bigl( N_{L_2\ l}{}^{L_1} \, \delta^{(4)}(m-s+M_2-M_1-S_2+S_1)
\nonumber \\
& & \qquad \qquad \qquad \qquad 
+ N_{L_2\ k-l}{}^{L_1}\, \delta^{(4)}(m-s+M_2-M_1-S_2+S_1+2)
\Bigr) \,.
\end{eqnarray}
Here $N_{L\ l}{}^{\hat{L}}$ denotes the level $k$ fusion rules of
$su(2)$, and $\chi_{(l,m,s)}$ is the character of the coset
representation. We note that the right hand side is invariant under
the field identification, $(l,m,s)\mapsto (k-l,m+k+2,s+2)$, as it must
be.  

\subsubsection{Topological spectrum}

It is now straightforward to read off from this formula the
topological spectrum between two such branes. If we consider only
D-branes of a fixed spin structure, we may set, without loss of
generality, $S=0$. Then we can restrict the open string sum without
loss of generality to the states with $s=0$; then the topological
states arise for $l=m$. 

\noindent For example, for the case $L=L_1=L_2$ (and $S_1=S_2=0$), the
number of topological states $T$ is 
\be\label{T1}
M_1=M_2 \quad ({\rm bosons}):\qquad 
|T| = \left\{
\begin{array}{cl}
L+2 \quad & \hbox{if $L$ is even} \\
L+1 \quad & \hbox{if $L$ is odd,}
\end{array} 
\right.
\ee
and 
\be\label{T2}
M_1=M_2 + 2 \quad ({\rm fermions}):\qquad 
|T| = \left\{
\begin{array}{cl}
L \quad & \hbox{if $L$ is even} \\
L+1 \quad & \hbox{if $L$ is odd.}
\end{array} 
\right.
\ee
Note that the boundary state corresponding to $L=k/2$ (which is an odd
number) has the same number of topological fermions and bosons in its
open string spectrum.

\subsection{The case $k/2$ even}

For $k/2$ even, the theory (\ref{neven}) has the $(k+2)$ Ishibashi 
states  
\be
|[l,0,s]\rangle\!\rangle \in \H_{[l,0,s]}\otimes \bar\H_{[l,0,s]} \,,
\ee
where $l$ and $s$ are both even, and the $(k+2)$ Ishibashi states 
\be
|[l,(k+2)/2,s]\rangle\!\rangle \in
\H_{[l,\frac{k+2}{2},s]}\otimes \bar\H_{[k-l,\frac{k+2}{2},s]} \,,
\ee
where $l$ is even and $s$ is odd. In addition, there are the two
Ishibashi states from the first sum in (\ref{neven})
\be
|[k/2,(k+2)/2,s\rangle\!\rangle \in 
\H_{[k/2,\frac{k+2}{2},s]}\otimes \bar\H_{[k/2,\frac{k+2}{2},s]} \,,
\qquad s=\pm 1 
\ee
and the two Ishibashi states from the second sum in (\ref{neven})
\be
|[k/2,0,s\rangle\!\rangle \in 
\H_{[k/2,0,s]}\otimes \bar\H_{[k-k/2,0,s]}\,, \qquad s=0,2 \,.  
\ee
In total there are therefore $2(k+4)$
Ishibashi states of the coset theory; they give rise to $(k+4)$
Ishibashi states of the $N=2$ theory with one spin structure
$\eta=+1$, and  $(k+4)$ Ishibashi states of the $N=2$ algebra with
the other $\eta=-1$. 

\noindent The corresponding boundary states are described by 
\begin{eqnarray}
|\!| L,M,S \rangle\!\rangle & = & \sqrt{\frac{2k+4}{2}} \, 
\sum_{l\, {\rm even}}
\Biggl( \sum_{s\  {\rm even}} 
\frac{S_{LMS,l0s}}{\sqrt{S_{000,l0s}}}\, |[l,0,s]\rangle\!\rangle  
 \nonumber \\
& & \qquad \qquad \qquad \qquad 
+  \sum_{s\ {\rm odd}} 
\frac{S_{LMS,l\frac{k+2}{2}s}}{\sqrt{S_{000,l\frac{k+2}{2}s}}}\, 
|[l,(k+2)/2,s]\rangle\!\rangle  \Biggr) \,. \label{boundneven}
\end{eqnarray}
Here $L+M+S$ is even, and $L\ne k/2$. We have the identifications 
\begin{eqnarray}
|\!|L,M,S\rangle\!\rangle & = & |\!| k-L,M+k+2,S+2 \rangle \!\rangle 
= |\!| k-L,M,S \rangle \!\rangle \nonumber \\
& = & |\!| L,M+4,S \rangle \!\rangle = 
|\!| L,M+2,S+2 \rangle \!\rangle \,. \label{idenne}
\end{eqnarray}
For fixed spin structure $\eta=+1$ (corresponding to $S=0,2$), we have
for every value of $L$ only two possible values of $M$, namely $M=0,2$
or $M=1,3$. Furthermore, because of the last identification, of the
four choices ($S=0,2$, $M=0,2$) or ($S=0,2$, $M=1,3$) only two
describe different boundary states. For $L\neq k/2$, the pair of
boundary states with  $L$ and $k-L$ account therefore each for two
different boundary states giving in total $k$ different boundary
states. The remaining four boundary states correspond to the `resolved'
boundary states for $L=k/2$, and will be described shortly. 

These boundary states satisfy the Cardy condition since their overlap
equals 
\begin{eqnarray}\label{nevenspec}
&& \langle\!\langle L_1,M_1,S_1 |\!|  
q^{L_0 + \bar{L}_0 - \frac{c}{12}} \,
|\!| L_2,M_2,S_2 \rangle \!\rangle 
=  \sum_{[l,m,s]} \, \chi_{(l,m,s)}(\tilde{q}) \, 
\delta^{(2)}(S_2+s-S_1) \times \nonumber \\
& & \qquad \qquad \times
\Bigl( N_{L_2\ l}{}^{L_1} +  N_{L_2\ k-l}{}^{L_1} \Bigr)
\, \delta^{(4)}(m-s+M_2-M_1-S_2+S_1) \,.
\end{eqnarray}
We observe again that the right hand side is invariant under the field
identification, $(l,m,s)\mapsto (k-l,m+k+2,s+2)$, as it must be. 

It is easy to see that the topological open string spectrum of these
branes is described by the same formulae as (\ref{T1}) and (\ref{T2})
above.

\subsubsection{The resolved branes}

The remaining boundary states are of the form
\begin{eqnarray}
|\!| k/2,M,S,\pm\rangle\!\rangle 
& = & \frac{1}{2} \,
|\!|k/2,M,S\rangle\!\rangle_{{\rm (\ref{boundneven})}}   \nonumber \\
& & \quad 
\pm \frac{\sqrt{2k+4}}{4} \, 
\Bigl( \sum_{s\ {\rm odd}} \, e^{i\pi M /2}\, e^{-i \pi s S/2}\,  
|[k/2,(k+2)/2,s]\rangle\!\rangle \nonumber \\
& & \qquad \qquad  \qquad \qquad 
+ \sum_{s\ {\rm even}} \, e^{-i \pi s S/2}\, 
|[k/2,0,s]\rangle\!\rangle \Bigr) \,.
\end{eqnarray}
Here $M+S$ is even, and we have the same identifications as in the 
second line of (\ref{idenne}). Thus these boundary states account for
four more boundary states (of a fixed spin structure).  

One easily checks that the overlaps of these boundary states with the
previously constructed boundary states is simply
\begin{eqnarray}\label{nevenrspec}
&& \langle\!\langle L_1,M_1,S_1 |\!|  
q^{L_0 + \bar{L}_0 - \frac{c}{12}} \,
|\!| k/2,M_2,S_2,\pm \rangle \!\rangle 
=  \sum_{[l,m,s]} \, \chi_{(l,m,s)}(\tilde{q}) \, 
\delta^{(2)}(S_2+s-S_1) \times \nonumber \\
& & \qquad \qquad \times
N_{k/2\ l}{}^{L_1} \, \delta^{(4)}(m-s+M_2-M_1-S_2+S_1) \,.
\end{eqnarray}
On the other hand, the overlaps involving two resolved branes 
is 
\begin{eqnarray}\label{nevenrespec}
&& \langle\!\langle k/2,M_1,S_1,\pm |\!|  
q^{L_0 + \bar{L}_0 - \frac{c}{12}} \,
|\!| k/2,M_2,S_2,\pm \rangle \!\rangle \\
& & \quad 
=  \sum_{[l,m,s]\ l=4n} \, \chi_{(l,m,s)}(\tilde{q}) \,  
\delta^{(2)}(S_2+s-S_1) \, 
\delta^{(4)}(m-s+M_2-M_1-S_2+S_1)\,,  \nonumber
\end{eqnarray}
and 
\begin{eqnarray}\label{nevenresspec}
&& \langle\!\langle k/2,M_1,S_1,\pm |\!|  
q^{L_0 + \bar{L}_0 - \frac{c}{12}} \,
|\!| k/2,M_2,S_2,\mp \rangle \!\rangle \\
& & \quad
=  \sum_{[l,m,s]\ l=4n+2} \, \chi_{(l,m,s)}(\tilde{q}) \,   
\delta^{(2)}(S_2+s-S_1) \, 
\delta^{(4)}(m-s+M_2-M_1-S_2+S_1)\,.  \nonumber
\end{eqnarray}
It is now straightforward to determine the topological open
string spectrum of these branes. As before, we may set, without loss
of generality (if we restrict ourselves to a specific spin structure)
$S_1=S_2=0$. Then the number of topological states on the $+$brane is
\begin{eqnarray}\label{T+}
& M_1=M_2 \quad ({\rm bosons}): \qquad & 
|T| = \frac{k}{4} + 1 \,, \\
& M_1=M_2 + 2 \quad ({\rm fermions}):  \qquad & 
|T| = 0  \,.
\end{eqnarray}
Obviously the result for the $-$brane is identical. On the other
hand, the topological spectrum between the $+$brane and the $-$brane
is  
\begin{eqnarray}\label{T+-}
& M_1=M_2 : \qquad & 
|T| = 0 \,, \\
& M_1=M_2 + 2 : \qquad & 
|T| = \frac{k}{4}  \,.
\end{eqnarray}

\subsection{The other GSO-projection}

As we mentioned before, we can also consider the 0A-like
GSO-projection. Then the spectrum of the D-model is for $k/2$ even
\be
\H = \bigoplus_{[l,m,s],\ l\ {\rm even}}
\Bigl( \left(\H_{[l,m,s]} \otimes \bar\H_{[l,m,-s]}\right)  
\oplus 
\left(\H_{[l,m,s]} \otimes \bar\H_{[k-l,m,-s]}\right) 
\Bigr) \,.
\ee
On the other hand, if $k/2$ is odd, the spectrum is 
\be
\H = \bigoplus_{[l,m,s],\ l\ {\rm even}}
\Bigl( \left(\H_{[l,m,s]} \otimes \bar\H_{[l,m,-s]}\right)   \Bigr) 
\, \oplus \,
\bigoplus_{[l,m,s],\ l\ {\rm odd}} 
\Bigl( 
\left(\H_{[l,m,s]} \otimes \bar\H_{[k-l,m,-s]}\right) 
\Bigr) \,.
\ee
In both cases we have the $k+2$ Ishibashi states  
\be
|[l,0,s]\rangle\!\rangle \in \H_{[l,0,s]}\otimes \bar\H_{[l,0,-s]} \,,
\ee
where $l$ and $s$ are both even, as well as the two Ishibashi states 
\be
|[k/2,0,s]\rangle\!\rangle \in 
\H_{[k/2,0,s]}\otimes \bar\H_{[k/2,0,-s]} \,,
\qquad k/2+s={\rm even} \,.
\ee
We therefore expect to have $k/2+2$ boundary states of each spin
structure. Some of these boundary states are given by 
\be\label{b11}
|\!| L,S \rangle\!\rangle = 
\sqrt{2k+4} \, \sum_{l,s\, {\rm even}}
\frac{S_{L0S,l0s}}{\sqrt{S_{000,l0s}}}\, |[l,0,s]\rangle\!\rangle  \,,
\ee
where $L=0,1,\ldots, k$ with $L\ne k/2$ and $S$ is defined modulo
$4$. We observe that  
\be\label{dboundeq}
|\!|L,S \rangle\!\rangle = |\!| k-L,S\rangle\!\rangle =
|\!|L,S+2\rangle\!\rangle \,.
\ee
For fixed spin structure (say $S$ even) there are thus $k/2$ such
boundary states. Their overlap equals 
\begin{eqnarray}
&& \langle\!\langle L_1,S_1 |\!|  
q^{L_0 + \bar{L}_0 - \frac{c}{12}} \,
|\!| L_2,S_2 \rangle \!\rangle \nonumber \\
& & \qquad \qquad \qquad 
=  \sum_{[l,m,s]} \, \chi_{(l,m,s)}(\tilde{q}) \, 
\delta^{(2)}(S_2+s-S_1) 
\Bigl( N_{L_2\ l}{}^{L_1} +  N_{L_2\ k-l}{}^{L_1} \Bigr) \,. \nonumber
\end{eqnarray}
Thus the open string spectrum on the brane $|\!|L,S\rangle\!\rangle$
has $2(L+1)$ topological states; the same is true for the open string
between the $|\!|L,S\rangle\!\rangle$ brane and its anti-brane. 

\noindent The remaining two boundary states are then 
\be
|\!| k/2,S,\pm \rangle\!\rangle = \frac{1}{2} 
|\!| k/2,S\rangle\!\rangle_{(\ref{b11})} 
\pm \sqrt{\frac{k+2}{4}} \, \sum_{s} e^{-i\pi S s/2}\, 
|[k/2,0,s]\rangle\!\rangle  \,,
\ee
where the sum over $s$ runs over $0,2$ if $k/2$ is even, and over 
$\pm 1$ if $k/2$ is odd.
For $k/2$ odd, these two branes are anti-branes of one another, 
{\it i.e.} 
$|\!|k/2,S,+\rangle\!\rangle = |\!|k/2,S+2,-\rangle\!\rangle$, and  
their overlaps equal
\begin{eqnarray}
&& \langle\!\langle k/2,S_1,\pm |\!|  
q^{L_0 + \bar{L}_0 - \frac{c}{12}} \,
|\!| k/2,S_2,\pm \rangle \!\rangle  \nonumber \\
&& \qquad \qquad \qquad
=  \sum_{[l,m,s], \, l\, {\rm even}} \, \chi_{(l,m,s)}(\tilde{q}) \, 
\delta^{(2)}(S_2+s-S_1) \, \delta^{(4)}(l + S_2 + s - S_1)
\,. \nonumber 
\end{eqnarray}
There are then $(k+2)/4$ topological states in the open
string spectrum of each of these branes, and $(k+2)/4$ topological
states in the open string spectrum between the brane and the
anti-brane. 

For $k/2$ even, on the other hand, each of the two branes
$|\!|k/2,S,\pm\rangle\!\rangle=|\!|k/2,S+2,\pm\rangle\!\rangle$ is its
own anti-brane. In this case their overlaps equal
\begin{eqnarray}
&& \langle\!\langle k/2,S_1,\pm |\!|  
q^{L_0 + \bar{L}_0 - \frac{c}{12}} \,
|\!| k/2,S_2,\pm \rangle \!\rangle  \nonumber \\
&& \qquad \qquad \qquad
=  \sum_{[l,m,s]} \, \chi_{(l,m,s)}(\tilde{q}) \, 
\delta^{(2)}(S_2+s-S_1) \, \delta^{(4)}(l) \,,
\nonumber \\[12pt]
&& \langle\!\langle k/2,S_1,\pm |\!|  
q^{L_0 + \bar{L}_0 - \frac{c}{12}} \,
|\!| k/2,S_2,\mp \rangle \!\rangle  \nonumber \\
&& \qquad \qquad \qquad 
=  \sum_{[l,m,s]} \, \chi_{(l,m,s)}(\tilde{q}) \, 
\delta^{(2)}(S_2+s-S_1) \, \delta^{(4)}(l + 2 )
\,. \nonumber 
\end{eqnarray}
In this case there are $k/4+1$ topological states in the open string
spectrum of either of these two branes, and $k/4$ topological states
in the open string between the two different branes.

\section{The matrix factorisation description}
\setcounter{equation}{0}

We now want to describe the matrix factorisations that correspond to
the above boundary states. As for the case of the A-type minimal
model, the two different GSO-projections correspond to two different 
superpotentials. We shall now discuss them in turn. 

\subsection{The first D-model} 

The theory whose boundary states we discussed in section~3.1 and 3.2
corresponds to the superpotential
\be
W_D = x^{n+1} - x y^2 \,,
\ee
where $n=k/2$. Its matrix factorisations have partially been studied
in \cite{KL3}, but, as we shall see, the list of factorisations given
there is incomplete. If we include another class of factorisations, we
obtain perfect agreement with the conformal field theory results
described above. Given that these boundary states generate all
boundary states of the superconformal field theory, we can turn the 
argument around and conclude that we have identified all fundamental
factorisations for this potential, {\it i.e.} that any factorisation
is equivalent to a direct sum of these fundamental factorisations.

\subsubsection{Rank 1 factorisations}

We start with a discussion of the rank 1 factorisations, all of which
have been discussed in \cite{KL3}. The most obvious factorisation is 
\be\label{simple}
{\cal R}_0: \qquad d_0=x^n-y^2, \quad d_1=x \,,
\ee
as well as its reverse. The bosonic spectrum of either factorisation
consists of two states, whereas the fermionic spectrum is empty. The
only boundary states with two bosons and no fermions in their open
string spectrum are the $L=0$ and $L=k$ boundary states of
(\ref{boundnodd}) and (\ref{boundneven}), respectively.
They must therefore correspond to the above factorisation (and its
reverse).   

Generically this is the only rank 1 factorisation, but if $n$ is even,
there are in addition the two resolved rank 1 factorisations, given by
\begin{eqnarray}
{\cal R}_+ & : & \qquad
 d_0 = (x^{n/2} + y)\,, \qquad d_1 = x ( x^{n/2} - y) \,, \nonumber \\
{\cal R}_- & : & \qquad
d_0 = x ( x^{n/2} + y) \,, \qquad d_1 = (x^{n/2} - y)\,, \nonumber 
\end{eqnarray}
as well as their reverses. Either of these factorisations has a purely
bosonic spectrum consisting of $n/2+1$ states, and there are $n/2$
fermionic operators propagating between ${\cal R}_+$ and 
${\cal R}_-$. Given that $n/2=k/4$, this matches precisely with the
spectrum of the resolved branes described in section~3.2.1 that also
only exist provided that $k/2$ is even.

\subsubsection{Rank 2 factorisations}

Next we turn to the rank $2$ factorisations. In \cite{KL3} the 
factorisations of the form 
\be
d_0= \left( \begin{array}{cc} x^l & \alpha \\ -\beta & -x^{n+1-l}
\end{array} 
\right) \,, \quad 
d_1= \left( \begin{array}{cc} x^{n+1-l} & \alpha \\ -\beta & -x^{l}
\end{array} 
\right)\,, \quad \alpha\beta=xy^2 
\ee
were considered. Exchanging $l$ with $n+1-l$ amounts to exchanging
$d_0$ and $d_1$, and hence to considering the reverse
factorisation. The same holds for the exchange of $\alpha$ and
$\beta$. The only two inequivalent choices for $\alpha,\beta$ are
therefore $\beta=x$ or $\beta=y$. As was shown in \cite{KL3} 
the choice $\beta=x$ leads to a class of factorisations that is
equivalent to the direct sum of the factorisation (\ref{simple}) and
its reverse. This leaves us with the factorisations
\be
{\cal S}_l: \qquad 
d_0= \left( \begin{array}{cc} x^l & xy \\ -y  & -x^{n+1-l}
\end{array} 
\right) \, \quad 
d_1= \left( \begin{array}{cc} x^{n+1-l} & xy \\ -y & -x^{l}
\end{array} 
\right)\,,
\ee
where $l=1,\ldots, n$. One easily finds that the factorisation 
${\cal S}_l$ has $2l$ bosons and $2l$ fermions. Comparing with the
conformal field theory analysis, one would then like to identify the 
factorisation ${\cal S}_l$ with the boundary state 
(\ref{boundnodd}) or (\ref{boundneven}) with $L=2l-1$. Note that 
$l\mapsto n+1-l$, which maps the factorisation to the reverse
factorisation, then corresponds to $L=2l-1 \mapsto k-L$, which maps
the brane to the anti-brane. [If $k/2$ is even, the anti-brane of 
$|\!|L,M,0\rangle\!\rangle $ is $|\!| k-L,M,2\rangle\!\rangle$, while
for $k/2$ odd it is $|\!| k-L,M,0\rangle\!\rangle$.]

As is clear from the conformal field theory analysis, the above
factorisations do not yet account for all branes since there are also
the boundary states (\ref{boundnodd}) or (\ref{boundneven}) with $L$
even. The corresponding class of factorisations can be directly
obtained from the factorisations (\ref{Afactor}) of the A-type minimal
model with potential $W_{A'}=x^n-y^2$: since  
$W_D=x(x^n-y^2)=xW_{A'}$, one obtains a factorisation for the D-model by
multiplying the matrix $d_1$ of any A-model factorisation by $x$. This
leads to the following factorisations  
\be\label{Dfactor}
{\cal T}_l : \qquad 
d_0^D= \left(\begin{array}{cc} x^l & y \\ -y & -x^{n-l} 
\end{array} \right)=d_0^{A'} \,, \quad
d_1^D= \left(\begin{array}{cc} x^{n-l+1} & xy \\ -xy & -x^{l+1} 
\end{array} \right) = x d_1^{A'} \,.
\ee
Here $l=0,\ldots,n$, and the factorisation ${\cal T}_l$ is equivalent
to ${\cal T}_{n-l}$, but neither is equivalent to ${\cal T}_l^r$. 
We now propose that the factorisation ${\cal T}_l$ corresponds to the
boundary state (\ref{boundnodd}) or (\ref{boundneven}) with $L=2l$. To
confirm this proposal, we have to determine the spectrum of these
factorisations. 

The calculation of the fermionic part of the spectrum proceeds exactly
as in the case of the A-model. For this, consider a fermion described
by its two components $(t_0,t_1)$. BRST-invariance means that  
\be
t_0\, d_1^D=-d_0^D\, t_1\,,
\ee
which can be solved by
\be
t_1=-\frac{1}{W_D}\, d_1^D \, t_0 \, d_1^D \,. 
\ee
This requires, of course, that the entries of the matrix appearing on
the right hand side are all divisible by $W_D$. In terms of the 
A-model data, this equation can be rewritten as 
\be
t_1=- x\, \frac{x}{W_D} \, d_1^{A'} \, t_0 \, d_1^{A'}
= - x\, \frac{1}{W_{A'}}\, d_1^{A'}\, t_0 \, d_1^{A'} \,,
\ee
where $W_D/x$ is the A-model potential $W_{A'}$. Without the
additional factor of $x$ on the right hand side, this condition is
simply the A-model result, implying a divisibility condition for the
entries of the matrix $d_1^{A'}\, t_0\, d_1^{A'}$. Since $x$ does not
divide $W_{A'}$ the divisibility conditions of the A- and D-model are 
therefore equivalent.   

\noindent Any solution for $t_0$ is BRST-trivial if
\be
t_0=-\phi_1 \, d_0^D + d_0 \, \phi_0^D \ .
\ee
Since $d_0^D=d_0^{A'}$ it is obvious that the fermionic part of the 
spectrum for the A- and D-model is the same for this class of
factorisations. Thus there are $2l$ fermions for the factorisation
${\cal T}_l$. 

Let us now turn to the bosonic spectrum. The condition for an operator
$(\phi_0,\phi_1)$ to be BRST invariant is 
\be
d_1^D \, \phi_1=\phi_0 \, d_1^D\,,
\ee
which is the same condition as in the A-model. Solving for $\phi_0$,
one obtains the divisibility condition  
\be
\phi_0^D = \frac{1}{W_D}\, d_1^D\, \phi_1\, d_0^D 
= \frac{1}{W_{A'}}\, d_1^{A'}\, \phi_1 \, d_0^{A'} \,,
\ee
which is explicitly (since ${\cal T}_l\equiv {\cal T}_{n-l}$ we may
assume that $l\leq n/2$) 
\be
y(\phi_1^{11} - \phi_1^{22}) + x^l \phi_1^{21} - x^{n-l} \phi_1^{12}
= 0 \qquad {\rm mod} \ W_{A'} \,,
\ee
where $\phi_1^{ij}$ denote the matrix elements of the $2\times 2$
matrix $\phi_1$. The possible solutions for $\phi_1$ are further
restricted by the requirement that the operator is BRST
non-trivial. BRST trivial operators can be written in the form (see
(\ref{BRSTex})) 
\ba
\phi_1^{11} &=& t_0^{11}\, x^{n-l+1} 
- t_0^{12}\, xy  + x^l \, t_1^{11} +y\, t_1^{21} \\ 
\phi_1^{12} &=& t_0^{11} xy - t_0^{12}\, x^{l+1} 
+ x^l\,  t_1^{12} + y\, t_1^{22} \\ 
\phi_1^{21} &=& t_0^{21} x^{n-l+1} 
- t_0^{22} \, x y  
- y\,  t_1^{11} - x^{n-l}\,  t_1^{21} \\
\phi_1^{22} &=& t_0^{21} x y - t_0^{22}\, x^{l+1} 
- y \, t_1^{12} - x^{n-l} \, t_1^{22} \,.
\ea
One can now use the freedom in $t_1$ to restrict the powers of $x$
appearing in the representatives of the BRST cohomology classes to lie
in the range $0,\dots, l-1$ (for $\phi^{11}_1$ and $\phi^{12}_1$)) and
$0,\dots, n-l-1$ (for $\phi^{21}_1$ and $\phi^{22}_1$)). On the other
hand, this then uses up the freedom described by $t_1$, and in
particular, the $y$-dependence of the solution cannot be eliminated
any longer. Indeed, in addition to the obvious $x$-dependent solutions
(that are as for the A-model, see (\ref{Abosons}))
\be
a_i=\left( \begin{array}{cc} x^i & 0 \\ 0 & x^i \end{array} \right)\,,
\quad 
a_{l+i}=\left( \begin{array}{cc} 0 & x^i \\ x^{n-2l+i} & 0 \end{array}
\right)\,,\quad  i=0,\dots , l-1 \,,
\ee
there are now the following two non-trivial solutions (for $\phi_1$)  
\be
b_1 = \left( \begin{array}{cc} y & 0 \\ 0 & y \end{array} \right)\,,
\qquad \hbox{and} \qquad 
b_2 = \left( \begin{array}{cc} 0 & y \\ yx^{n-2l} & 0 \end{array}
\right)  \,.
\ee
This means, in particular, that there are two more bosons than
fermions for each factorisation ${\cal T}_l$. This is then in perfect
agreement with the conformal field theory spectra (\ref{T1}) and
(\ref{T2}) for the branes with $L=2l$.  
\smallskip

\noindent In summary, we therefore propose the identifications:
$$
\begin{array}{|l|c|l|}
\hline
{\rm Boundary}\,\, {\rm state} & {\rm Matrix}\,\,{\rm factorisation} 
& \\ \hline \hline
|\!|0,M,S\rangle\!\rangle  & {\cal R}_0 \,, {\cal R}_0^r & \\
\hline
|\!|2l-1,M,S\rangle\!\rangle & {\cal S}_l\,, {\cal S}_l^r &
l=1,\ldots, [(n+1)/2]\\
\hline
|\!|2l,M,S\rangle\!\rangle  &  {\cal T}_l\,, {\cal T}_l^r &
l=0,\ldots, [n/2]  \\
\hline
|\!| k/2,M,S,\pm\rangle\!\rangle & 
{\cal R}_{\pm}\,, {\cal R}_{\pm}^r & n \,\, {\rm even} \\
\hline
\end{array}
$$
As a consistency check we note that we have now identified the 
boundary state with $L=0$ with two factorisations, namely with 
${\cal R}_0$ and with ${\cal T}_0$. Thus we need to have 
that these two factorisations are actually equivalent (in the 
sense of (\ref{equiv})), and this is indeed easily
checked. Furthermore, for $k/2=n$ even, the boundary state 
$|\!|k/2,M,S\rangle\!\rangle$ is not fundamental, but can be resolved as
explained in section~3.2.1. We have identified these resolved branes
with the factorisations ${\cal R}_\pm$; thus we expect that for $n$
even the factorisation ${\cal T}_{n/2}$ is equivalent to the direct
sum of ${\cal R}_+$ and ${\cal R}_-$, and this is also
straightforwardly confirmed.  

Another consistency check concerns the various flows of D-brane
configurations. By switching on suitable tachyons, one can show
that there are flows from the point of view of the matrix
factorisation (such flows were first discussed, for the case of the
A-model, in \cite{HLL2}) 
\begin{eqnarray}
{\cal R}_0 \oplus {\cal R}_0^r & \rightarrow & {\cal S}_1 \nonumber \\
{\cal S}_l \oplus {\cal R}_0 & \rightarrow & {\cal T}_l \nonumber \\
{\cal T}_l \oplus {\cal R}_0^r & \rightarrow & {\cal S}_{l+1} \,. 
\nonumber
\end{eqnarray}
These flows are easily seen to be compatible with the RR-charges of
the corresponding boundary states. They are also in agreement with
what one expects based on the analysis of \cite{Fred}. Finally, we
note that the $L=0$-brane (together with the two resolved branes if
$n$ is even) generates all D-brane charges since all D-branes (except
for the resolved branes) can be obtained from it by the above
flows. The corresponding matrix factorisations of these three D-branes
all have rank $1$. This is similar to the situation for the tensor
product of two A-models where the permutation branes (whose
factorisations also have rank $1$) generate all the charges
\cite{BG1}.

\subsection{The other D-model}

Finally we consider the superpotential 
\be
W_{D'}=x^{n+1}-xy^2 - z^2
\ee
that should correspond, following the logic of \cite{KL3}, to the
D-model with the other GSO-projection that was discussed in
section~3.3. The easiest class of factorisations of this model are of
tensor product type: for any factorisation of $W_D$ one obtains a 
factorisation of $W_{D'}$, where one factorises $z^2$ as
$z^2=zz$. Since the spectrum on a brane with superpotential $W=z^2$
consists of the  identity $\phi_0=\phi_1=1$ and a fermion
$t_0=-t_1=1$, it is very simple to calculate the spectrum of these
factorisations. The number of bosons is simply given by the sum of the
number of bosons and fermions of the corresponding factorisation of
the D-model $W_D$. The same holds for the  number of fermions. In
particular, the number of bosons always equals the number of fermions
for any factorisation of this type.  

We now propose that we can identify the boundary states (\ref{b11})
with these tensor product factorisations. As is explained in
section~3.3, each of these branes has $2(L+1)$ topological states in
its open string spectrum (irrespective of whether $L$ is even or odd),
and the same is true for the open string between the brane and its 
anti-brane. This then matches precisely with the above spectrum of
the corresponding tensor product factorisations, given our previous
identifications of the boundary states with the matrix factorisations
for the $W_D$ model: in particular, the sum of the number of bosons
and fermions in (\ref{T1}) and (\ref{T2}) is precisely equal to
$2(L+1)$ for all $L$.   

The analysis works similarly for the two resolved branes 
$|\!|k/2,S,\pm\rangle\!\rangle$ for $k/2$ even. This leaves us with
identifying the resolved brane $|\!|k/2,S,\pm\rangle\!\rangle$ for
$n=k/2$ odd, which does not come from such a tensor factorisation. 
For $n$ odd there are however additional factorisations of the form
\be\label{addi}
d_0=\left( \begin{array}{cc} x^{\frac{n+1}{2}}-z & \alpha \\
   -\beta & -(x^{\frac{n+1}{2}}+z) \end{array} \right)\,, \quad
d_1=\left( \begin{array}{cc} x^{\frac{n+1}{2}}+z & \alpha \\
   -\beta & -(x^{\frac{n+1}{2}}-z) \end{array} \right)\,, \quad 
\alpha\beta=xy^2 \,.
\ee
Exchanging $\alpha$ and $\beta$ maps the factorisation to its reverse,
and thus effectively the only inequivalent choices are $\beta=x$ and
$\beta=y$. As before for the case of the D-model, one may expect that
$\beta=x$ is in some sense trivial, and indeed one can show that this
factorisation is equivalent to the tensor product factorisation
corresponding to the D-model factorisation ${\cal R}_0$,   
\be
d_0=\left( \begin{array}{cc} x^{n}-y^2 & z \\
   -z & -x \end{array} \right)\,, \quad
d_1=\left( \begin{array}{cc} x & z \\
   -z & -(x^{n}-y^2) \end{array} \right)\,. \quad 
\ee
On the other hand, the topological spectrum of the factorisation 
(\ref{addi}) with $\beta=y$ has $(n+1)/2$ bosonic and fermionic
states, which thus agrees with the result of section~3.3. 

We have therefore also managed to identify the boundary states of this
D-model with the factorisations of the superpotential $W_{D'}$. Since
we know that the boundary states we have described generate all
possible boundary states, the same must be true for the above
factorisations.

\section{Matrix factorisations and singularity theory}
\setcounter{equation}{0}

Matrix factorisations are also a well-known tool in the theory
of singularities in mathematics. In the case of complex dimension 
two, the simple singularities are known to have an ADE classification.
The relevant singularities are described by  hypersurfaces in $\BC^3$
characterised by the equations\footnote{By a change of variables one
can also rewrite the first equation as $W=x^n-y^2-z^2=0$.}
\be\label{hyper}
\begin{array}{rrl}
A_{n-1}: &W=x^n - yz =& 0 \\ 
D_{n+2}    : &W=x^{n+1}-xy^2 -z^2 =& 0 \\ 
E_6 : &W=x^4+y^3+z^2=& 0 \\ 
E_7 : &W=yx^3 + y^3 + z^2 =& 0 \\ 
E_8 : & W=x^5 + y^3 + z^2 =& 0\,.
\end{array}
\ee
The resolution of the singularities is obtained by blow-ups, where
each node of the associated ADE Dynkin diagram corresponds to an
exceptional divisor. For the case of these simple surface
singularities, the explicit description of the blow-up can be encoded
in a matrix factorisation of the polynomial $W$ defining the
hypersurface \cite{eise,yosh,GV}.

D-branes on such singular geometries have recently been  discussed in
\cite{orlov}. There it was proposed to define a category $D_{Sg}$ that
is supposed to capture the topological sector of B-type branes on
singular spaces. On smooth  manifolds, any coherent sheaf has a finite
resolution of locally free sheaves of finite type. On singular spaces
this is no longer the case and Orlov defined $D_{Sg}$ to be the
quotient of the bounded derived category of coherent sheaves modulo
those sheaves that have such a finite resolution. It is shown in
\cite{orlov} that this category is  equivalent to the category of
matrix factorisations; in particular, any object in $D_{Sg}$
corresponds to a matrix factorisation, and the open string spectrum
that is described in $D_{Sg}$ in terms of morphisms of modules
describing the branes, corresponds exactly to the BRST invariant
spectrum of the Landau Ginzburg theory. More precisely, consider a
Landau Ginzburg potential $W: \BC^{n} \to \BC$ with an isolated
critical point at the origin. [In our case the relevant LG potentials
are $W_{ADE}:\BC^3\to \BC$, and thus describe precisely the singular
hupersurfaces of (\ref{hyper}).] Denoting the fiber of $W$ over $0$ by
$S_0$, \cite{orlov} allows us to establish a relation between
$D_{Sg}(S_0)$ and the Landau-Ginzburg category. For this, we associate
to any factorisation $W=d_0d_1$  
$$
\Bigl(
\xymatrix{
P_1 \ar@<0.6ex>[r]^{d_1} &P_0 \ar@<0.6ex>[l]^{d_0}}
\Bigl)
$$
the short exact sequence
\begin{equation}\label{shseq}
0\lto P_1 \stackrel{d_1}{\lto} P_0 \lto \Coker d_1 \lto 0\,.
\end{equation}
The geometrical object associated to the factorisation is then the 
sheaf $\Coker d_1$, which, since it is annihilated by $W$, is a sheaf
on $S_0$. For further details on this functor and mathematical proofs,
we refer to \cite{orlov}. 

\subsection{Singular geometry vs. Landau Ginzburg model}

As we have just seen, the very same matrix factorisations that 
describe B-type D-branes in the Landau Ginzburg theory with
superpotential $W$ also characterise the B-type D-branes of the
singular hypersurface $W=0$ and its possible resolutions.
Despite their formal similarities, these two description are 
however rather different: string theory on the singular geometry is,
in sigma-model language, a singular limit of a theory with $c=6$ that
has no well-behaved conformal field theory description. On the other
hand, the Landau Ginzburg model flows to a perfectly well defined
conformal field theory, namely the $N=2$ minimal model with
$c=3k/(k+2)$.   

{}From the closed string point of view the relation between singular
geometries and $N=2$ conformal field theories has been studied before
in \cite{VW,Warn,LVW,Mart}. In particular, characteristic properties
of the Landau Ginzburg model, such as the central charge and the
chiral ring, have been compared with concepts appearing in singularity
theory (singularity index, local ring of $W$). The agreement of the 
matrix factorisations thus provides a natural extension of this
correspondence to the open string sector. Since the matrix
factorisations determine (the topological part of the) open string
spectrum, this may even suggest that the B-type D-branes of the  
Landau-Ginzburg model (or of the $N=2$ minimal model)
provide a worldsheet description of the D-branes
that become massless in the singular geometry.

\subsection{Singular geometry vs. the orbifold description}

String theory compactified on a singular surface does not have a
well-defined perturbation theory since at the singular point
non-perturbative (D-brane) states become massless. However, it is
possible to give a well-defined conformal field theory description for
a closely related background where a non-zero $B$-field has been
switched on for all singular cycles. (This $B$-field prevents the
D-branes from becoming massless in the singular limit, and therefore
avoids the breakdown of string perturbation theory \cite{aspI,aspII}.)
The relevant conformal field theory is the orbifold theory
$\BC^2/\Gamma$, where $\Gamma$ is a finite subgroup of $SU(2)$.
Orbifold theories are well behaved, and it is known how to describe
their D-branes. In particular, following \cite{DM}, the different
D-branes are in essence characterised by the representation of the
orbifold group $\Gamma$ that acts on the corresponding Chan-Paton
indices. The charges are then generated by the branes that are
associated to the non-trivial irreducible representations of $\Gamma$.
If we associate to each such representation a node of a (Dynkin)
diagram, and connect nodes if the open string between the
corresponding D-branes has a (massless) hypermultiplet in its
spectrum, we recover precisely again the corresponding ADE Dynkin
diagrams. Furthermore, the dimension of the irreducible representation
of $\Gamma$ equals the Kac-label of the corresponding node. This
suggests, in particular, that the orbifold description captures at
least some of the structure of the geometry described by $W=0$. 

Since the fundamental D-branes of the orbifold theory also give rise
to the same ADE Dynkin diagram, they should be in natural one-to-one 
correspondence with the matrix factorisations of the LG
potential.\footnote{However, the open string spectra should not (and
do not) agree: because of the resolution certain massless open string 
states of the Landau Ginzburg theory become massive in the orbifold
theory.} In
fact this relation can be understood fairly directly. As described in 
\cite{curto} following the ideas of \cite{yosh,GV}, for each
irreducible representation of $\Gamma$ one can find a  module of the
$\Gamma$-invariant part of $\BC[X,Y]$. Unless the 
representation of $\Gamma$ is trivial, these modules are not free, 
and the relations among the generators of the module
can be described by a matrix. This is then precisely the matrix
that appears as one factor of the matrix factorisation. For the two
cases of interest in this paper, this can be done explicitly as
follows.  

\subsubsection{A-type singularity}

In the case of A-type singularities, the generator of the orbifold
group acts on $\BC^2$ as 
\be
g X= \xi X\,, \qquad gY = \xi^{-1} Y\,, 
\qquad \xi = e^{\frac{2\pi i}{n}} \,.
\ee
The singular surface can be described by the polynomial subring in the
two variables $X,Y$ that is invariant under the orbifold action. This
invariant subring is generated by $x=XY$, $y=Y^n$ and $z=X^n$. These
polynomials are however not independent, but satisfy the equation
$x^n-yz=0$, which is just the singular hypersurface equation of the
$A$-type singularity. 

The orbifold group $\Gamma$ is in this case $\BZ_n$, which has 
$n-1$ irreducible non-trivial representations. All of these
representations are one-dimensional: for $l=1,\ldots,n-1$ 
the corresponding representation associates to the generating element
$g\in\Gamma$ the phase $e^{\frac{2\pi i l}{n}}$. 

These $n-1$ representations are then in one-to-one correspondence to
the matrix factorisations of the Landau-Ginzburg potential
(\ref{ASP}), or equivalently, the boundary states of the minimal model
(\ref{single}), whose label take values 
$L=l-1=0,\dots, k=n-2$.\footnote{The superpotential $W=x^n-yz$ (that is
equivalent by a change of variables to $W=x^n-y^2-z^2$) differs from
$W_A=x^n$ by changing the GSO-projection twice. One would therefore
expect that these two superpotentials are equivalent. From the point
of view of singularity theory this equivalence is known as Kn\"orrer
periodicity \cite{Knoerrer}.}  
As explained in \cite{curto} we can associate to each 
representation of $\Gamma$ a module for the invariant part of the
polynomial ring in two variables. For example, for the representation
labelled by $l$ the corresponding module is generated by $X^l$
and $Y^{n-l}$. This is not a free module since we have the relations
between the generators $s_1=Y^{n-l}$, $s_2=X^l$:
\be
x^l s_1 - y s_2 =0\,, \qquad -z s_1 + x^{n-l}s_2 =0 \,.
\ee
The matrix of relations
\be
d_1=\left( \begin{array}{cc} x^l & -y  \\ -z & x^{n-l} \end{array}
\right) 
\ee
is then one of the matrices appearing in the matrix
factorisation. Conversely, we can apply the functor of Orlov and
recover the module (and thus the representation of $\Gamma$) from the
factorisation.

\subsubsection{ D-type singularity}

The D-type minimal model corresponds to the case where the orbifold is
a binary dihedral group. Its two generators act on $\BC^2$ as
\be
g=\left( \begin{array}{cc} \beta & 0 \\ 0 & \beta^{-1} \end{array} 
\right)\,,
\qquad
h=\left( \begin{array}{cc} 0 & 1 \\ -1 & 0 \end{array} \right)\,, 
\qquad {\rm where} \ \beta=e^{\frac{\pi i}{n}} 
\ee
and satisfy the relations
\be
g^{2n} = {\bf 1}\,, \qquad 
h^2 = g^n \,, \qquad 
hgh^{-1}=g^{-1} \,.
\ee
The ring of invariant polynomials under the group is generated by
\be
x=(XY)^2\,, \qquad y=\frac{1}{2}(X^{2n}+Y^{2n})\,, 
\qquad z=\frac{i}{2}(XY)(X^{2n}-Y^{2n})\,,
\ee
which satisfy the D-type hypersurface equation $x^{n+1}-xy^2-z^2=0$. 
Generically, the irreducible representations of this group are
two-dimensional, with 
\be
\rho^{(l)}(g) = \left( \begin{array}{cc} \beta^l & 0 \\ 0 & \beta^{-l}
\end{array} \right)
\ee
and
\be
\rho^{(l)}(h) = \left( \begin{array}{cc} 0 & 1 \\ -1 & 0 \end{array}
\right) 
\quad (l \ {\rm odd}) \,, \qquad
\rho^{(l)}(h) = \left( \begin{array}{cc} 0 & i \\ -i & 0 \end{array}
\right) 
\quad (l \ {\rm even}) \,.
\ee
Here $l=0,1,\ldots, 2n-1$, and the representations 
$\rho^{(l)}$ and $\rho^{(2n-l)}$ are equivalent. Furthermore, the 
representations $\rho^{(0)}$ and $\rho^{(n)}$ are not irreducible
but can be further decomposed into one-dimensional representations:  
$\rho^{(0)}$ contains the trivial representation and the one where 
$g\to 1, h\to -1$. Likewise, $\rho^{(n)}$ can be decomposed into
the two one-dimensional irreducible representations
$g \to -1$ and $h\to \pm 1$ ($l$ even) or $h\to \pm i$ ($l$ odd).  

These representations are exactly in one-to-one correspondence with
the boundary states of section~3.3, or equivalently the matrix
factorisations of section~4.2. More explicitly, the representation
$\rho^{(l)}$ corresponds precisely to the boundary state with 
$L=l$. Since $n=k/2$, the reducible representation $\rho^{(n)}$
corresponds then to the non-fundamental brane with $L=k/2$ which can
be resolved into two boundary states (that correspond in turn to the
two one-dimensional representations with $g\mapsto -1$). Also the 
identification $\rho^{(l)}\simeq \rho^{(2n-l)}$ mirrors the
equivalence of boundary states (\ref{dboundeq}). As in the A-case, one
can also relate these representations to certain modules of the
invariant algebra $\BC[X,Y]^\Gamma$, and obtain the corresponding
matrix factorisations in this manner (see \cite{curto}). It is maybe
remarkable that the blow-ups that correspond to the one-dimensional
representations of $\Gamma$ are quite different to those that are
associated to the two-dimensional representations. 
\bigskip

For completeness we should also mention yet another class
of models with $c=6$ describing the motion of strings on ALE
spaces. In this approach (which is due to \cite{oogvaf}), the minimal
models are tensored with another theory that contributes the missing
central charge to get $c=6$. To be more precise, \cite{oogvaf}
consider the tensor product of two coset theories  
$SL(2,\Real)/U(1) \times SU(2)/U(1)$, and orbifold by an appropriate
discrete group to impose the charge integrality condition. In terms of
Landau-Ginzburg potentials, these models can be written as
\be
\begin{array}{rrl}
A_{n-1}: &W=-\mu t^{-n} + x^n - yz =& 0 \\ 
D_{n/2+1}:  & W=-\mu t^{-n} + x^{n/2}-xy^2 -z^2 =& 0 \,,
\end{array}
\ee
which are again subject to an integer charge projection.
As argued in \cite{oogvaf}, the value of the $B$-field in these models
is $0$, but now $\mu\ne 0$, and we are therefore discussing the
resolved geometry. The D-branes of this model have been investigated
in \cite{Lerc,LLS}. If one wants to consider B-type D-branes in these
models, one can effectively reduce the discussion to A-type D-branes
in the minimal model part. The reason is that the orbifold procedure
(followed by the mirror map) maps A-type to B-type branes, and, as
noticed in \cite{Lerc,LLS}, the contribution of the non-compact coset
can be captured by universal factors. Since this description
involves effectively the A-type D-branes of the minimal model, the 
relation to matrix factorisations and branes in the singular geometry
is however less evident.

\bigskip

\centerline{\large \bf Acknowledgements}
\vskip .2cm

This research has been  partially supported by a TH-grant 
from ETH Zurich, the Swiss National Science Foundation and the Marie
Curie network `Constituents, Fundamental Forces and Symmetries of the
Universe' (MRTN-CT-2004-005104). We thank Stefan Fredenhagen, Wolfgang
Lerche and Sebastiano Rossi for useful discussions.

\end{document}